\documentclass[a4paper,12pt]{article}
\pdfoutput=1 
\usepackage{jheppub} 
\usepackage[T1]{fontenc} 
\usepackage{amsfonts,amssymb,amsmath}
\usepackage{graphicx,color,float}
\usepackage{wrapfig}
\usepackage{subcaption}

\newcommand{\be}{\begin{equation}}
\newcommand{\ee}{\end{equation}}
\newcommand{\bea}{\begin{eqnarray}}
\newcommand{\eea}{\end{eqnarray}}
\newcommand{\beal}{\begin{aligned}}
\newcommand{\eeal}{\end{aligned}}

\usepackage{multicol}
\usepackage[framemethod=TikZ]{mdframed}
\usepackage{lipsum}

\definecolor{palatinate}{RGB}{128, 49, 123}

\mdfdefinestyle{MyFrame}{
linecolor=blue,
outerlinewidth=2pt,
roundcorner=20pt,
innertopmargin=\baselineskip,
innerbottommargin=\baselineskip,
innerrightmargin=20pt,
innerleftmargin=20pt,
backgroundcolor=gray!20!white}

\title{Higgs Vacuum Decay in a Braneworld}

\author[a]{Leopoldo Cuspinera}
\author[a,b,c]{Ruth Gregory}
\author[d]{Katie M.\ Marshall}
\author[d]{Ian G.\ Moss}

\affiliation[a]{Institute for Particle Physics Phenomenology, Department of Physics, 
Durham University, South Road, Durham DH1 3LE, UK}
\affiliation[b]{Department of Mathematical Sciences, Durham University,
South Road, Durham, DH1 3LE, UK}
\affiliation[c]{Perimeter Institute, 31 Caroline Street North, Waterloo, 
ON, N2L 2Y5, Canada}
\affiliation[d]{School of Mathematics, Statistics and Physics, Newcastle University, 
Newcastle Upon Tyne, NE1 7RU, UK}

\emailAdd{j.l.cuspinera@durham.ac.uk}
\emailAdd{r.a.w.gregory@durham.ac.uk}
\emailAdd{k.marshall6@newcastle.ac.uk}
\emailAdd{ian.moss@newcastle.ac.uk}

\abstract{
We examine the effect of large extra dimensions on 
vacuum decay in the Randall-Sundrum (RS) braneworld paradigm.
We assume the scalar field is confined to the brane, and 
compute the probability for forming an ``anti de Sitter'' (AdS)
bubble inside a critical flat RS brane.
We present the first full numerical solutions for the brane instanton 
considering two test potentials for the scalar field. We explore 
the geometrical impact of thin and thick bubble walls, and compute the
instanton action in a range of cases. We conclude
by commenting on a more physically realistic potential relevant for the
standard model Higgs. For bubbles with large backreaction, the extra
dimension has a dramatic effect on the tunnelling rate, however, for
the weakly backreacting bubbles more relevant for realistic Standard 
Model potentials, the extra dimension has little impact.
}

\preprint{DCPT-19/19}

\begin{document}

\maketitle

\section{Introduction}

When Coleman and de Luccia \cite{CDL} pioneered the study of vacuum 
decay in curved spacetime, they described the possibility as `the ultimate 
ecological catastrophe'. Whilst the comment was somewhat tongue in cheek, 
the recent measurement of the Higgs mass \cite{ATLAS:2012ae,Chatrchyan:2012tx} 
and the realisation that the standard model Higgs field could well be in a 
metastable state \cite{Krive:1976sg,1982Natur.298,Sher:1988mj,
Isidori:2001bm,Degrassi:2012ry,Gorsky:2014una,Bezrukov:2014ina,
Ellis:2015dha,Blum:2015rpa}, has brought the catastrophe a little closer to reality!
Fortunately, the timescale for decay according to Coleman et al.\ 
\cite{coleman1977,callan1977,CDL}, (see also \cite{Kobzarev:1974cp}), is 
sufficiently large that we would seem not be troubled, except that the Coleman
results are computed in a highly symmetric background. Recent work by
two of us \cite{GMW,BGM1,BGM2,BGM3,Gregory:2016xix} has
argued that taking into account inhomogeneities such as primordial 
black holes can dramatically shorten the lifetime of the false vacuum
(see also \cite{PhysRevD.35.1161,Berezin:1987ea} for early work, and
\cite{Tetradis:2016vqb,Chen:2017suz,Gorbunov:2017fhq,
Mukaida:2017bgd} for alternate perspectives). 

Apart from primordial black holes, the other possible scenario in which 
small black holes might occur is in particle collisions if there are large
extra dimensions. Large extra dimension scenarios were introduced 
initially to provide an alternate, geometric, resolution of the hierarchy
problem. The idea that our four dimensional Planck scale is 
derived from a higher dimensional Planck mass close to the standard 
model scale \cite{ArkaniHamed:1998rs,Antoniadis:1998ig,Randall:1999ee,
Randall:1999vf}; we then live on a four-dimensional {\it brane} embedded
in a higher dimensional spacetime. In such scenarios it is easier to form 
black holes in particle collisions (see e.g.\ \cite{Giddings:2001bu,
Dimopoulos:2001hw,Landsberg:2003br}). 
Our relatively high Planck scale, $M_p^2 = 1/8\pi G_N$, is then the result of a 
geometric hierarchy coming from an integration over the extra dimensions.

In a previous paper \cite{Cuspinera:2018woe}, we computed the 
probability for seeded decay with a brane black hole, following the 
notion that small black holes could also occur in particle collisions 
if there are large extra dimensions.  As in the straightforward four 
dimensional (4D) case, we found the decay rate to be significantly 
enhanced over the Hawking evaporation rate for a range of small
mass black holes. However, we did not compare the seeded nucleation
process to an unseeded brane Coleman de Luccia (CDL) equivalent
rate, hence we could not clarify the extent to which enhancement of tunnelling
was due to the black hole, or the extra dimensions. In this work, we
address this question, and explore the vacuum decay of a brane-scalar
in the absence of any black hole. The instantons we will consider will
be the true brane equivalents of the CDL solution.
Early work on brane instantons \cite{Gregory:2001xu,Gregory:2001dn} 
focussed largely on constructing the Euclidean solutions and presented
results on the action within the thin wall approximation, subsequent work 
either focussed on compact instantons in, or near, the thin wall limit 
\cite{Davis:2005hf}, or approximate Hawking-Moss type instantons
\cite{Demetrian:2005sr} and bulk scalar instantons \cite{Dudas:2007hq}.
See also some interesting ideas on RS brane decay by 5D
``bubble of nothing'' type processes \cite{Ida:2001qw,Ochiai:2001fh}, 
as well as instantons in DGP \cite{Izumi:2007gs,Sbisa:2014gwh}.

In this paper we consider vacuum decay of a scalar field 
localised on a brane embedded in a five-dimensional anti-de Sitter (AdS) bulk. 
We first review the derivation of the instanton equations in \S \ref{sec:instantoneq},
then present numerical solutions for the scalar CDL-equivalent brane
instanton in \S \ref{sec:setting}. In \S \ref{sec:computation-action} we 
turn to a computation of the action, showing how to renormalise the 
instanton action properly, and computing the action for a range of 
potentials and Planck mass hierarchies before concluding in \S \ref{sec:concl}.

\section{The Instanton Equations of Motion}
\label{sec:instantoneq}

In the Randall-Sundrum (RS) model, spacetime is 5 dimensional with
a negative cosmological constant living in the bulk. This negative 
curvature of spacetime causes a localisation of the graviton on the brane,
the background solution being a brane with energy and tension
equal and precisely tuned to the bulk cosmological constant, 
giving a flat brane at $z=0$:
\be
ds^2 = e^{-2|z|/\ell} \eta_{\mu\nu} dx^\mu dx^\nu - dz^2
\ee
where  $\ell^2 = -6/\Lambda_5$ is the AdS curvature scale.
The local negative curvature of the bulk supports the brane 
tension $\sigma$ that is easily calculated from the Israel junction 
conditions \cite{Israel:1966}:
\be
{\cal K}^{(+)}_{\mu\nu} = -\frac1\ell \eta_{\mu\nu} \quad \Rightarrow \qquad
{\cal K}^+_{\mu\nu} -  {\cal K}^+ \eta_{\mu\nu} = \frac3\ell \eta_{\mu\nu}
= 4\pi G_5\sigma \eta_{\mu\nu}
\ee
One can add energy momentum to the brane, for example a 
``brane cosmological constant'', so that $\sigma$ is greater or less
than the critical value \cite{Chamblin:1999ya,Kaloper:1999sm,
Kraus:1999it,Karch:2000gx,Binetruy:1999ut,BCG,Maartens:2000fg}, 
a cosmological fluid, or a perturbative localised source. In all cases, 
the intuitive visualisation of brane matter is that it causes the braneworld to
bend as first pointed out by Garriga and Tanaka \cite{Garriga:1999yh}
(see also \cite{Shiromizu:1999wj,Sasaki:1999mi}).

We are interested here in pure false vacuum decay, i.e.\ the brane 
equivalent of a Coleman de Luccia (CDL) instanton that 
is a Euclidean solution to the Einstein-plus-brane-scalar field equations 
that has $O(4)$ symmetry on the brane.
This level of symmetry is mathematically equivalent to a cosmological 
braneworld solution: there is a brane coordinate $\tau$ upon which
the brane solution depends, and a coordinate that tracks the warping
in the bulk. If we assume that the full brane plus bulk solution also has 
$O(4)$ symmetry, then a ``generalised Birkhoff theorem'' applies
\cite{BCG}, and the bulk equations
of motion can be fully integrated with the brane now
following a trajectory in the bulk consistent with the local 
energy-momentum of the instanton solution (for proof see
\cite{BCG,Gregory:2001xu,Gregory:2001dn}).

To find these equations of motion, we take a simple scalar field 
lagrangian on the brane: 
\be
{\cal L}_\phi=\frac12g^{\mu\nu}\phi_{,\mu}\phi_{,\nu}+V(\phi).
\ee
The general bulk admitting an $O(4)$ symmetric brane solution is a 
Schwarzschild-AdS black hole \cite{BCG,Gregory:2001xu}, 
however, as we are computing the brane equivalent
of the CDL instanton, we will take the pure AdS$_{5}$ 
spacetime in the bulk
\be
ds_{\text{bulk}}^{2} = h(r) dt^2 + \frac{dr^2}{h(r)} + r^2 d\Omega^{2}_{I\!I\!I}\;,
\quad h(r) = 1+ \frac{r^2}{\ell^2}
\label{adsbulk}
\ee
as a bulk black hole induces a cosmological radiation source on the brane
\cite{Binetruy:1999ut,Maartens:2000fg,BCG}.

The brane traces out a submanifold in \eqref{adsbulk} that can be
parametrised by intrinsic coordinates $\{\tau, \theta^\alpha\}$ ($\alpha=1,2,3$):
\be
X^{\mu}=(t(\tau),a(\tau),\theta^\alpha)
\ee
where $\tau$ is chosen to be the proper time parameter on the brane
\be
h \dot{t}^2+ \frac{\dot{a}^2}{h} = 1,
\label{taudef}
\ee
so that the induced brane metric is identical to the CDL geometry:
\be
ds_{\text{brane}}^2 = d\tau^2 + a^2(\tau) d\Omega^{2}_{I\!I\!I}\;.
\ee
The scalar field depends only on $\tau$, and the energy-momentum is
readily found to be
\be
\beal
T_{\tau\tau} &= \sigma + V - {1\over 2} \dot\phi^{2}
=  \frac{3 {\cal E}}{4\pi G_{5}} \\
T_{\alpha\beta} &=[\sigma + V + {1\over 2} \dot\phi^{2}] g_{\alpha\beta} 
=\frac{3{\cal T}}{4\pi G_5}g_{\alpha\beta}
\eeal
\label{EandT}
\ee
that sources the brane trajectory.

The Israel junction equations are then
\be
\beal
K^+_{\tau\tau} = \frac1{h\dot{t}} \left( \ddot a - \frac{h'(r)}{2} \right)
&= 2{\cal E} - 3 {\cal T} \\
K^+_{\alpha\beta} = - \frac{\dot{t}\, h }{a} g_{\alpha\beta}
&= - {\cal E} g_{\alpha\beta}\;,
\eeal
\ee
usually expressed in the cosmological format of Friedmann
and conservation of energy momentum equations:
\be
\beal
\left ( \frac{\dot a}{a} \right)^2 &= \frac{1}{a^2} + \frac{1}{\ell^2} - {\cal E}^2\\
0&= \dot{\cal E} + \frac{3\dot{a}}{a} ({\cal E} - {\cal T})
\eeal
\ee

For numerical integration of the scalar field, it is more useful to use
the Raychaudhuri equation, and substituting in the form of the energy-momentum
\eqref{EandT} we finally arrive at the full set of brane-scalar instanton equations:
\be
\beal
\left ( \frac{\dot a}{a} \right)^2 &= \frac{1}{a^2}  - \frac{8\pi G_N }{3}
\left( V- \frac12 \dot \phi^{2} \right)
- \left( \frac{4\pi G_N\ell}{3}\right)^2  \left( V- \frac12\dot \phi^{2} \right)^{2}\\
\frac{\ddot{a}}{a} &=  - \frac{8\pi G_N }{3} \left( V + \dot \phi^{2} \right)
- \left( \frac{4\pi G_N\ell}{3}\right)^2  
\left( V- \frac12 \dot{\phi}^2 \right) \left( V + \frac52 \dot{\phi}^2 \right)\\
\ddot{\phi} + \frac{3\dot{a}}{a} \dot{\phi} &=\frac{\partial V}{\partial \phi} \;.
\eeal
\label{eq:EOM}
\ee
where we have substituted the Newton constant $G_N = G_5/\ell$
in the gravitational coupling. These are precisely the SMS \cite{Shiromizu:1999wj}
(Shiromizu-Maeda-Sasaki) equations with vanishing Weyl term, also
analysed in \cite{Demetrian:2005sr} for the Hawking-Moss case. 
As $\ell$ drops, gravity becomes more
strongly localised on the brane, hence the 4D limit is $\ell\to0$, and
\eqref{eq:EOM} become the 4D instanton equations.

It is also worth noting that the critical RS brane (with $V=\dot{\phi}=0$)
has $\dot{a} \equiv 1$. This leads to the brane trajectory
\be
r = a(\tau) = \tau \;, \qquad 
t(\tau) = \frac{\ell}{2} \log (1+ \tau^2/\ell^2)
\label{RSbrane}
\ee
in terms of the original coordinates \eqref{adsbulk}. This is a less
familiar form for the critical RS brane, obtained because we are
solving for the brane in bulk global coordinates, rather than the usual
Poincare patch. The trajectory can easily be transformed to its
familiar form using
\be
e^{z/\ell} = \frac{e^{t/\ell}}{\sqrt{1+r^2/\ell^2}}\;\;,
\qquad x^i = e^{z/\ell} r n^i_4
\ee
where $n_4$ is the unit vector in 4 dimensions.

\section{The Scalar Brane Instanton}
\label{sec:setting}

In order to investigate vacuum decay, we use two basic model scalar potentials.
The first is a standard quartic potential $V_q$, with a potential barrier between 
a false and true vacuum. It is convenient to parametrise this potential with 
the value of $\phi=\phi_M$ at the maximum and $\phi=\phi_V$
at the global minimum:
\be
V_q(\phi) = g\left [ \frac{\phi^4}{4}  - \frac{\phi^3}{3} \left (\phi_V+\phi_M\right)
+ \frac{\phi^2}{2}\phi_V\phi_M \right]
\label{quadpot}
\ee
The potential vanishes at the false vacuum $\phi=0$ and the value at the true 
vacuum is
\begin{equation}
V_q(\phi_V) = \frac{g}{12}\,\phi_V^3(2\phi_M-\phi_V).
\end{equation}
Note that since we require $V_q(\phi_V)<0$, $\phi_V>2\phi_M$.

The second potential we wish to investigate, $V_h$, more closely 
approximates the Higgs potential. 
The form of this potential has one local minimum and a barrier, 
where on the far side the potential does not turn up again until it 
reaches very high field values. This allows for the possibility of a 
phase transition and the nucleation of a true vacuum bubble. 
The potential takes the form
\be
V_h(\phi)=\frac14\lambda_{\rm eff}(\phi)\phi^4.
\ee
where the effective coupling
\be
\lambda_{\rm eff}=g\left\{\left(\ln{\frac{\phi}{M_p}}\right)^4
-\left(\ln{\frac{\Lambda}{M_p}}\right)^4\right\}
\ee
$g\sim 10^{-5}$ is a constant that can be used to tune to the potential to 
closely fit the standard model Higgs potential. 

\begin{figure}[h]
\centering
\includegraphics[width=0.47\textwidth]{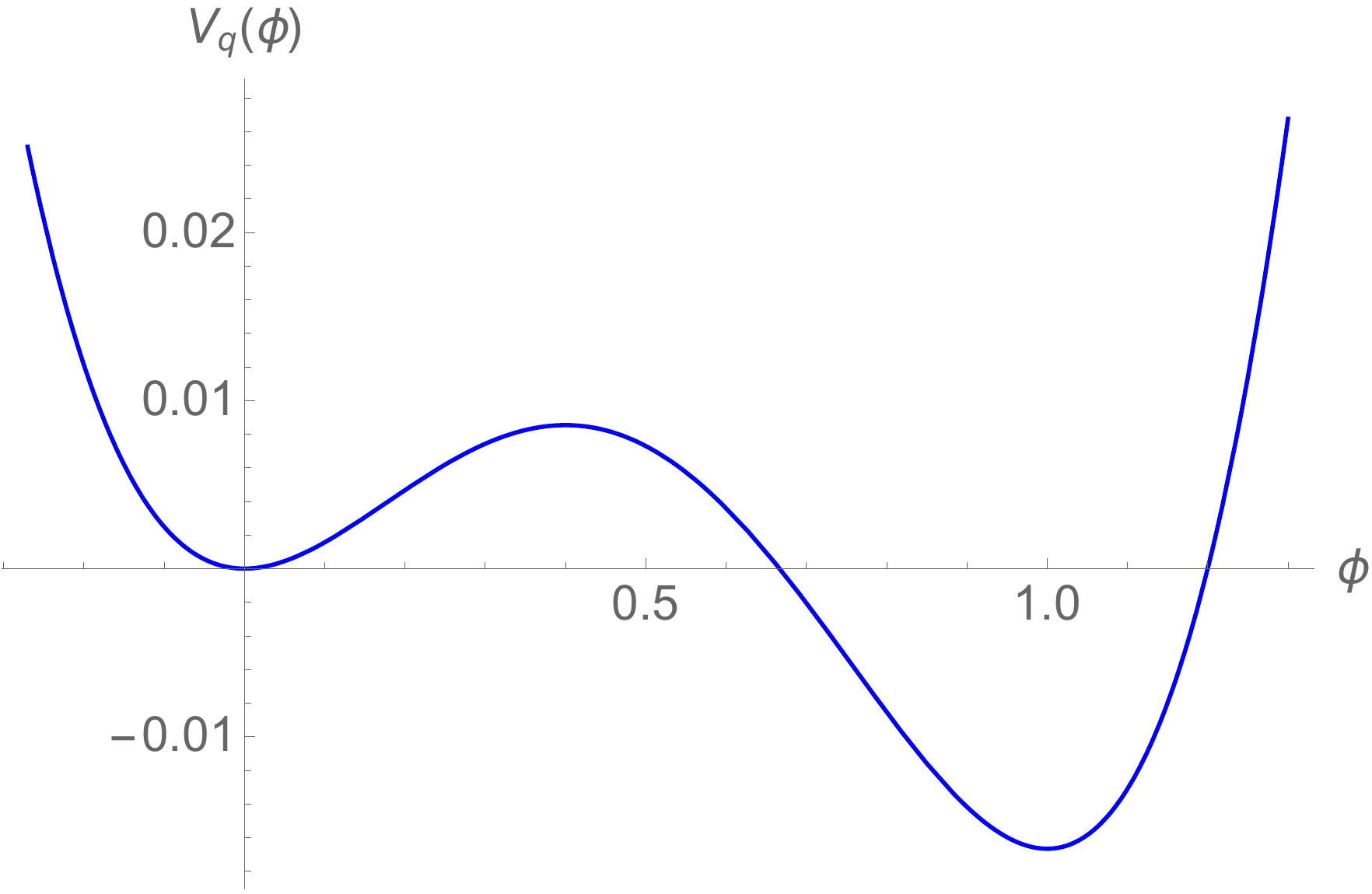}\qquad
\includegraphics[width=0.47\textwidth]{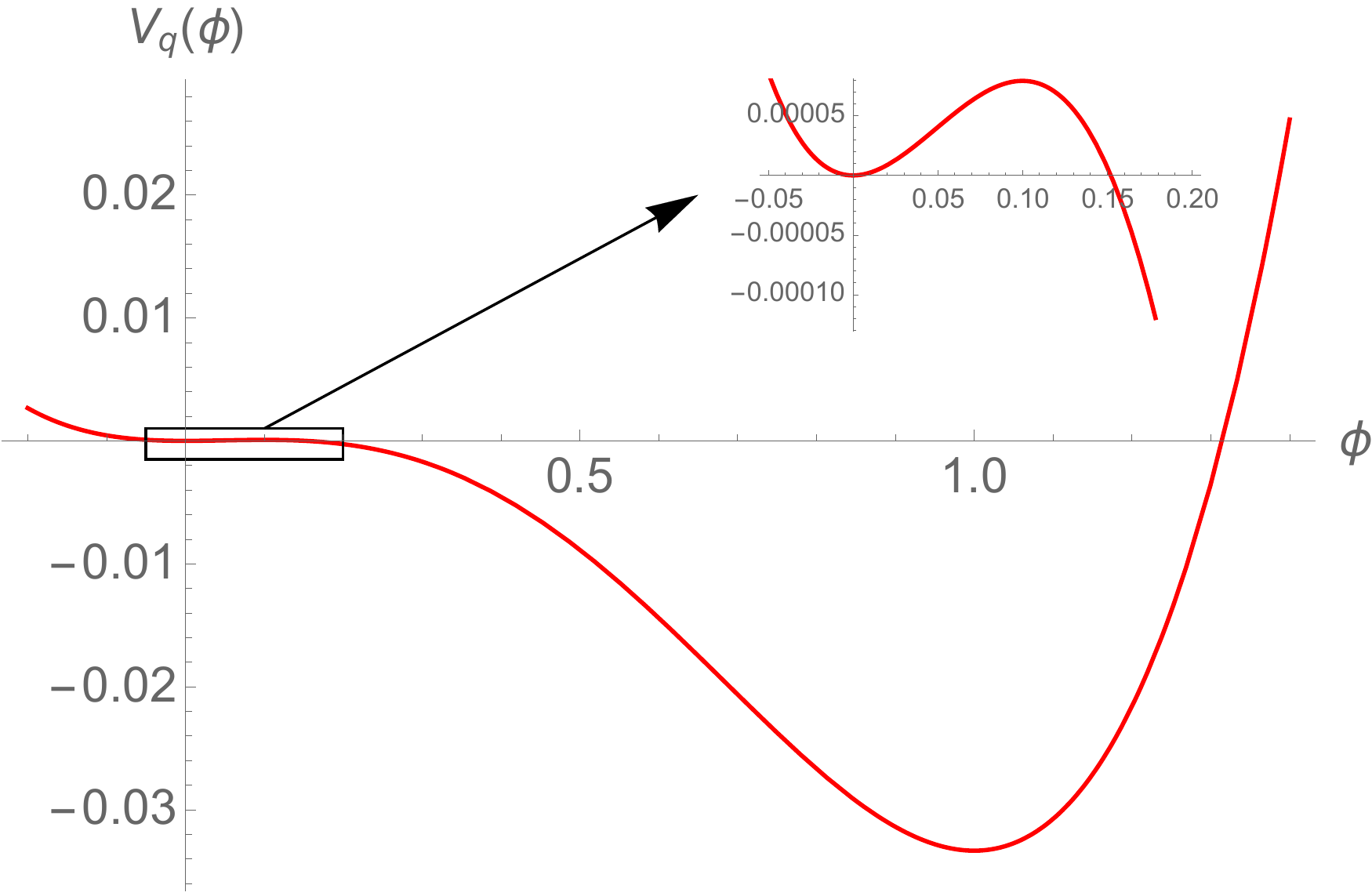}\\
\caption{The $V_q$ potentials referred to in the text. 
On the left in blue with $\phi_M=0.4$,
and $\phi_V=1$ (with $M_p=1$), corresponding to a well-defined bubble
wall. On the right in red the potential more closely approximated the Higgs
potential, with $\phi_M = 0.1$, and corresponds to a thick wall bubble.}
\label{fig:vq}
\end{figure}

In each case, we integrate \eqref{eq:EOM} from the centre of
the instanton, $\tau=0$, looking for a solution that asymptotes
the flat critical RS trajectory \eqref{RSbrane}. However, note that
because we set boundary conditions at $\tau=0$ of
$a=0$, $\dot a=1$ and $\dot\phi=0$, the flat geometry at
large $\tau$ is $\phi \to \phi_{\rm fv}$, $a \to \tau + c$ --
integrating through the bubble wall produces an offset in the 
value of $r$ relative to $t$. While this is not particularly relevant
to the form of the bubble solution, for which $a(\tau)$ is important,
it is a crucial observation for the computation of the action, as
we will return to in the next section.

The quadratic potential \eqref{quadpot} is particularly useful for exploring
the variation from thin to thick bubble walls, and for varying backreaction
strengths. To illustrate this, we present results for two representative
potentials, one giving a strongly backreacting thin wall, with parameter values 
$g=1, \phi_V=M_p, \phi_M = 0.4M_p$, and the other a weakly backreacting
thick wall with parameter values $g=1/2, \phi_V=M_p, \phi_M = 0.1M_p$;
in both cases the Planck scales are $M_5 = 0.4$, $M_p=1$, hence the bulk
AdS lengthscale is $\ell = 1/M_5^3 = 125/8$. Figure \ref{fig:vq} shows
the potential $V_q$ for these two choices of parameters; note the thin
wall potential (shown in blue) has a significant potential barrier between
the vacua, but less well represents a Higgs-type potential, whereas the 
thick wall potential (shown in red) more closely resembles the Higgs
potental, having a very small barrier relative to the global minimum.

\begin{figure}[h]
\centering
\includegraphics[width=0.47\textwidth]{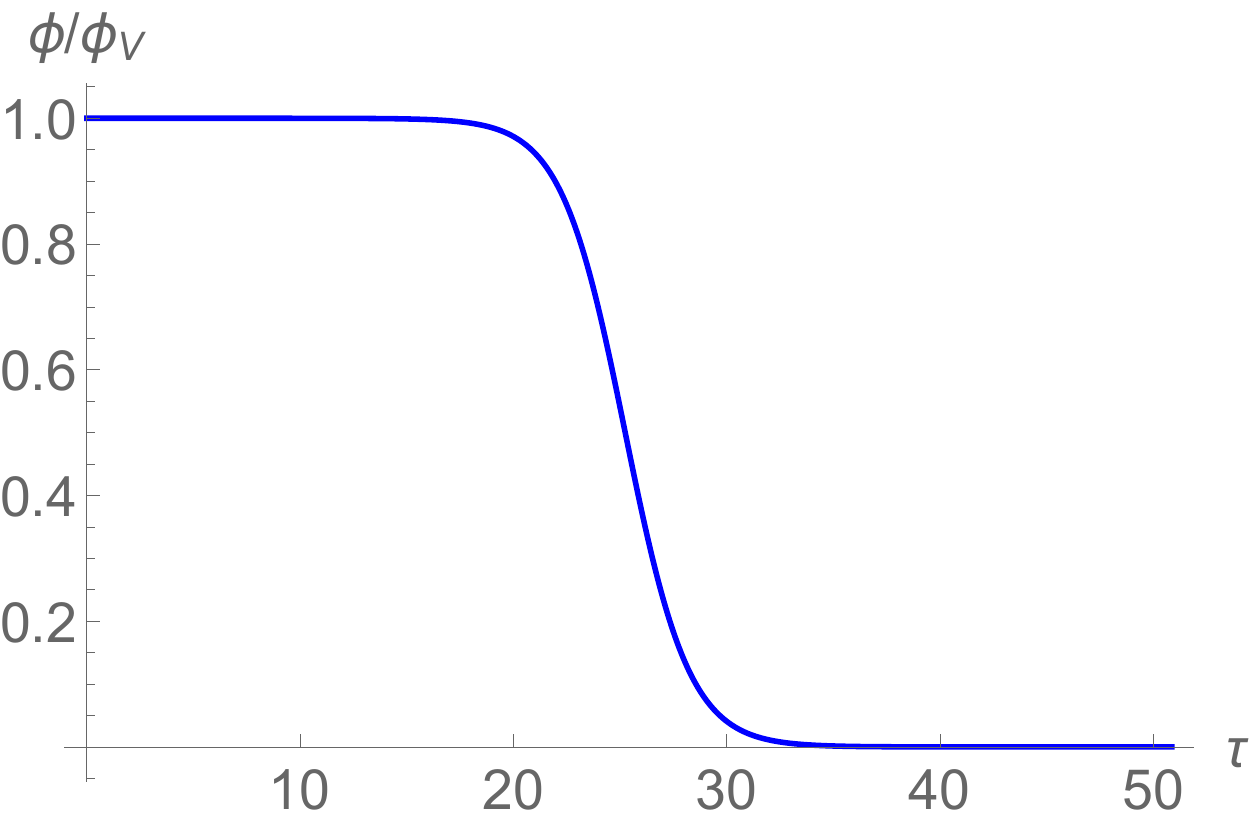}\quad
\includegraphics[width=0.47\textwidth]{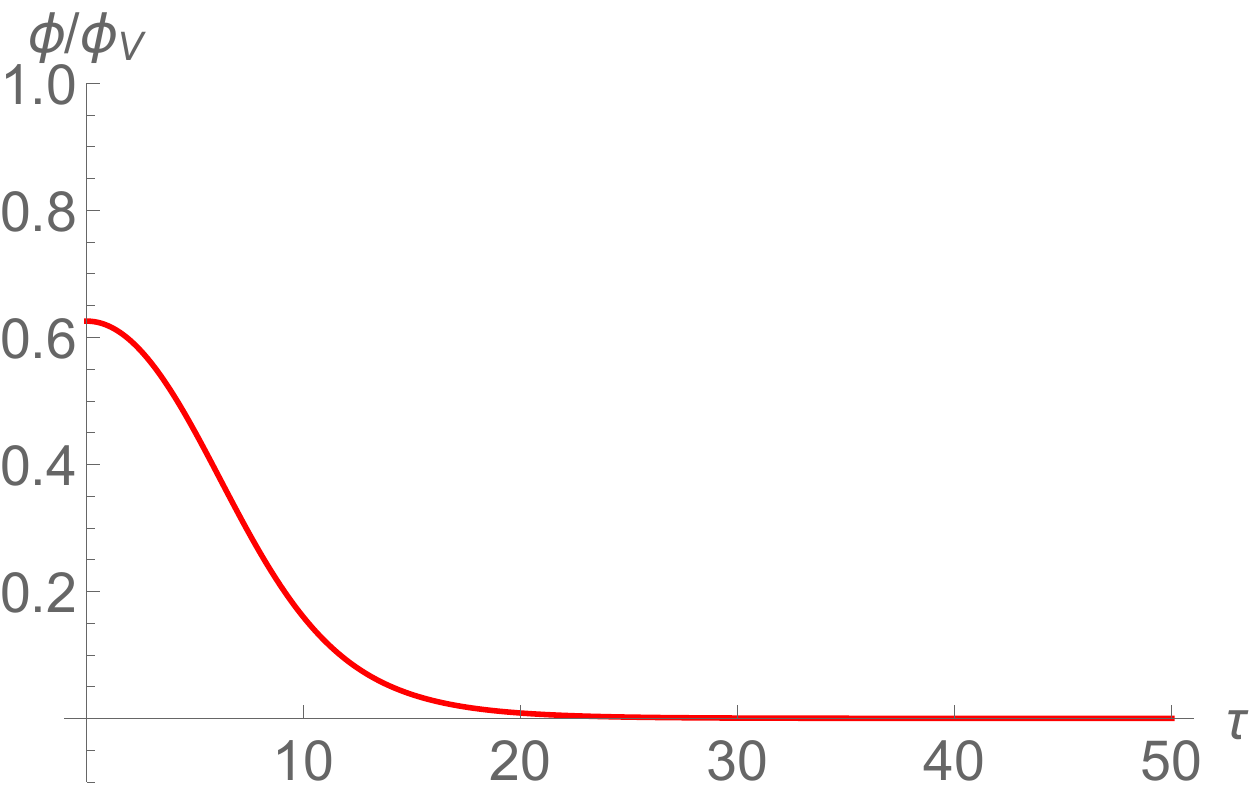}
\caption{The scalar field solution for the potentials shown in figure \ref{fig:vq}.
Once again, blue corresponds to the thin wall bubble, here clearly seen as a step
in $\phi$, and red to the thick wall bubble.}
\label{fig:phiplot}
\end{figure}

\begin{figure}
\centering
\includegraphics[width=0.3\textwidth]{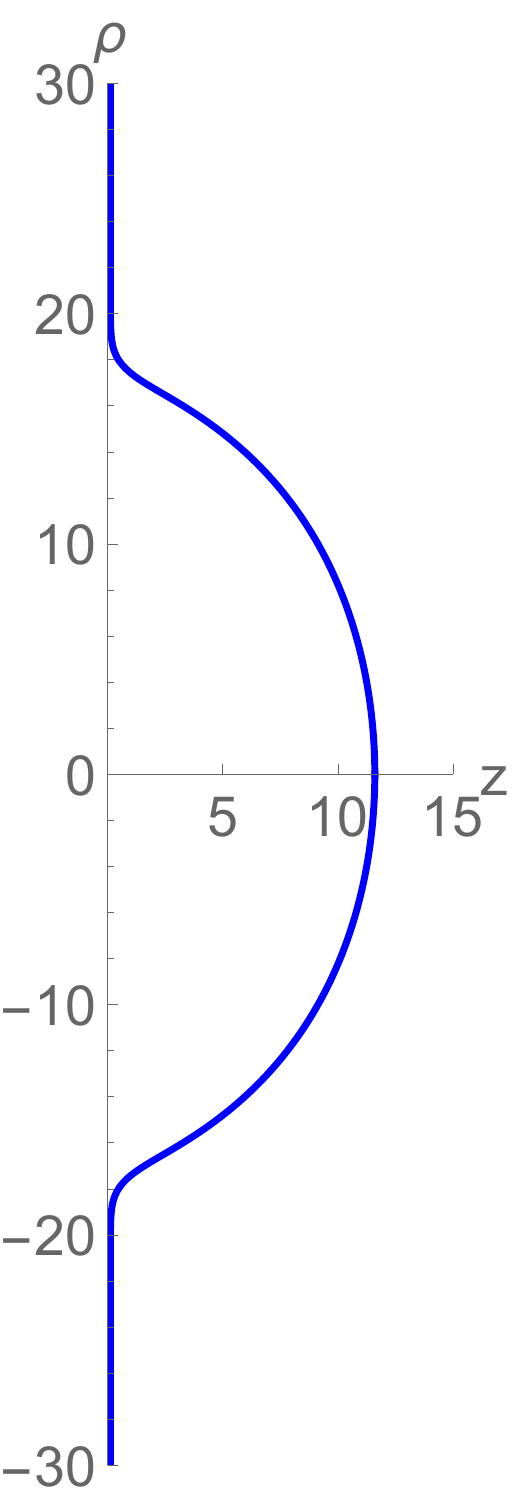}\qquad\qquad
\includegraphics[width=0.3\textwidth]{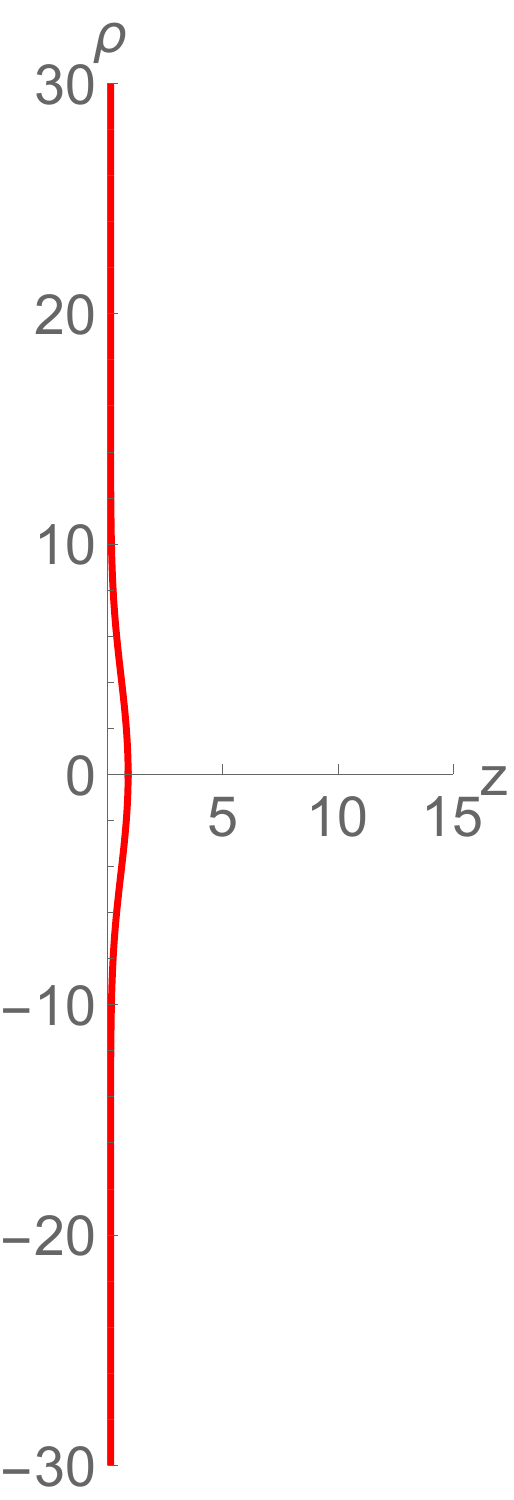}
\caption{The geometry of the brane with bubble embedding shown in Poincare 
coordinates, as is usual for the flat RS brane. (Colour scheme as fig.\ \ref{fig:vq}.)}
\label{fig:braneplot}
\end{figure}

The scalar field solution is shown in \ref{fig:phiplot}, and demonstrates clearly
the distinction between the potentials: the thin wall has a clear,
sharp transition from false to true vacuum around $\tau \sim 25$, whereas
the thick wall does not even reach the true vacuum by the centre of the bubble.
The effect of the bubble on the embedding of the brane is shown in figure
\ref{fig:braneplot}. The strongly backreacting thin wall brane shows the 
transition between the flat RS critical asymptotic false vacuum brane, and the 
sub-critical true vacuum AdS embedding in the interior of the brane. The
weakly interacting thick wall has a much less significant displacement, and
does not reach the spherical shape of the sub-critical brane.

\section{Computation of the action}
\label{sec:computation-action}

Having found the Euclidean brane bubble solutions, we now need
to compute their action, in order to find the leading order exponential
behaviour of the tunnelling probability. We first derive the action
for a general brane scalar solution, given a large $r$ cutoff, then
discuss the proper background subtraction.
The Euclidean action is given by 
\be
\beal
S &=  \frac{1}{8\pi G_{5} } \int_{M^+}d^5x\left(-R_5+2\Lambda_5\right) 
+ \int_{\partial M^+} d^4x \, \left [ \frac{2K}{8\pi G_{5}} +
\frac{1}{2}(\partial \phi)^2 +V + \sigma \right],\\
&= \int_{M^+}\frac{d^5x }{\pi G_{5} \ell^2}
+ \int_{\partial M^+}\!\!\!d^4x \, \left [ 
\frac{\dot{\phi}^2}{6} -\frac13(V + \sigma) \right]
\eeal
\label{eq:Action}
\ee
that is formally infinite for the background false vacuum critical brane solution. 
Note that this action is written in the Gibbons-Hawking boundary format,
with the brane being a boundary of a 5D manifold, the normal $n_{\mu}$ 
pointing in to the manifold -- this expression manifestly includes the 
$\mathbb{Z}_2$ symmetry of the brane. 

In order to find the instanton action, we first apply a cutoff well outside the
radius of the bubble. We define the cutoff by $a(\tau_R) = R$, and bound the
bulk coordinates by $r\leq R$, and $t \leq t_b(\tau_R)$, where $t_b$ is the
value of $t$ on the brane found by integrating the relation \eqref{toftau} below.
Note that $t$ is also bounded below by $t_b(0)$. We obtain:
\be
S_R = \frac{2\pi}{G_{5} \ell^2} \int_{t_b(0)}^{t_b(R)} dt \,\int_0^R dr\, r^3
+ 2\pi^2 \int_0^{\tau_R} d\tau a^3(\tau) \left [ 
\frac{\dot{\phi}^2}{6} -\frac13(V + \sigma) \right]
\ee
Now, whereas the bulk integral is naturally expressed in terms of the bulk
coordinates $t$ and $r$, the brane integral and the instanton solution are
naturally expressed in terms of the intrinsic coordinate $\tau$. While 
we can easily identify $r=a(\tau)$, the relation to the bulk time coordinate is
differential:
\be
\frac{dt}{d\tau} = \frac{{\cal E}a(\tau)}{1+a^2/\ell^2}\,.
\label{toftau}
\ee
Using this relation, we can rearrange the bulk integral, integrating first
with respect to $r$, then translating to a $\tau$ integral to finally obtain
\be
S_R = \frac{\pi^2}{3} \int_0^{\tau_R} d\tau \frac{a^3}{1+a^2/\ell^2} 
\left [ \dot{\phi}^2 -2V -2 \sigma \right]
\label{actionnonren}
\ee
\begin{wrapfigure}{r}{0.5\textwidth}
\centering
\includegraphics[width=0.45\textwidth]{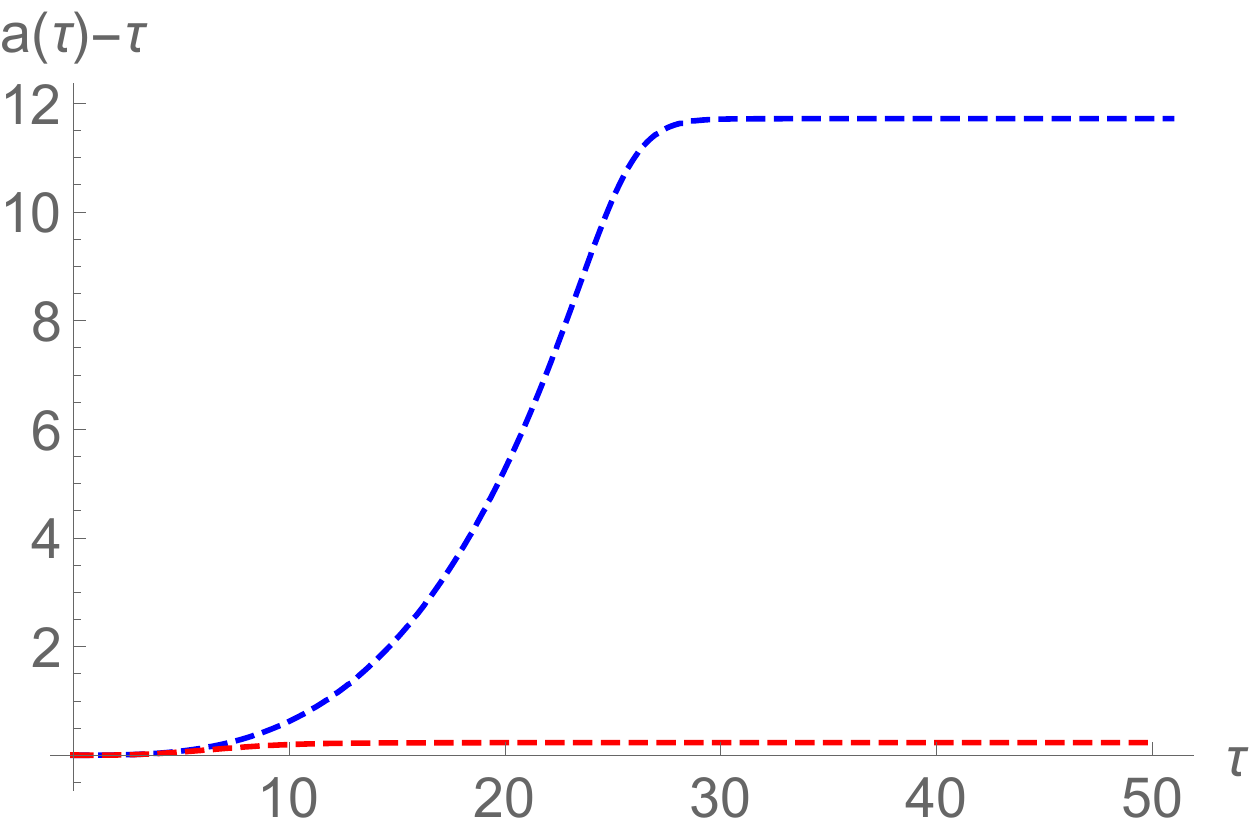}
\caption{The offset between $a$ and $\tau$ for the thin and thick wall,
with colour scheme as fig.\ \ref{fig:vq}.}
\label{fig:tauoffset}
\end{wrapfigure}
where $\tau_R$ is defined as $a(\tau_R)=R$. This integral now is in a 
simple ``brane'' format, and we can easily insert in the solutions of the
scalar instanton equations. The integral diverges as $\tau_R^2$ 
for large $R\sim \tau_R$, however, outside the bubble, both instanton and
false vacuum branes are identical, thus once we subtract the background
false vacuum action this divergence will be removed. 

To subtract the background false vacuum a crucial observation is that
the false vacuum action is \emph{not} obtained simply by deleting all
but the $\sigma$ term in the above \eqref{actionnonren}, since
not only is $a(\tau)$ different, but also the value of $\tau$ at which the 
brane radius becomes equal to $R$ (see figure \ref{fig:tauoffset}). 
We must therefore perform one
final manipulation to get the instanton action. The critical false vacuum 
brane action is
\be
S_{FV} = \frac{-2\pi^2}{3} \int_0^{\tau'_R} 
\frac{a^3(\tau')\sigma d\tau' }{1+a^2(\tau')/\ell^2}
= \frac{-2\pi^2}{3} \int_0^R  \frac{a^3\sigma da}{1+a^2/\ell^2} 
\ee
but now that this is expressed as an integral over $a$, we can compare this
to the $a$ integral for the bubble:
\be
S_{\text{bub}} = \frac{\pi^2}{3} \int_0^R \frac{da}{\dot{a}}
\frac{a^3}{1+a^2/\ell^2} 
\left [ \dot{\phi}^2 -2V -2 \sigma \right]
\ee
and using this expression gives the final subtracted action for the instanton
as
\be
\beal
B = S_R - S_{FV} &= \frac{2\pi^2}{3} \int_0^R \frac{da}{\dot{a}}
\frac{a^3}{1+a^2/\ell^2} 
\left [ \frac{\dot{\phi}^2}{2} - V + (\dot{a}-1) \sigma \right]\\
&= \frac{2\pi^2}{3} \int_0^{\tau_R} d\tau
\frac{a^3}{1+a^2/\ell^2} 
\left [ \frac{\dot{\phi}^2}{2} - V + (\dot{a}-1) \sigma \right]
\eeal
\ee
now expressed as an integral over the brane time-coordinate (and
numerical integration parameter) $\tau$.
\begin{figure}
\includegraphics[width=0.5\linewidth]{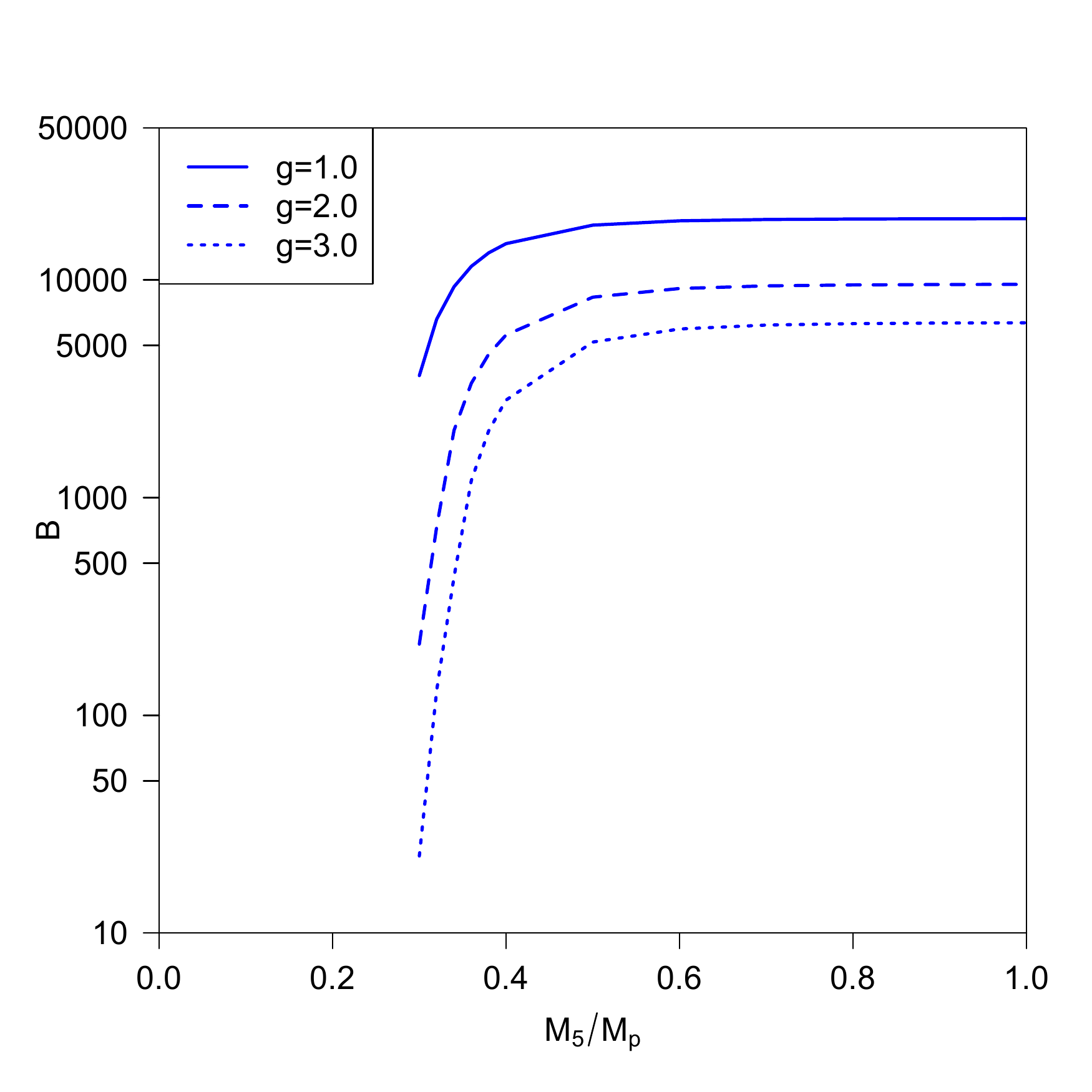}
\includegraphics[width=0.5\linewidth]{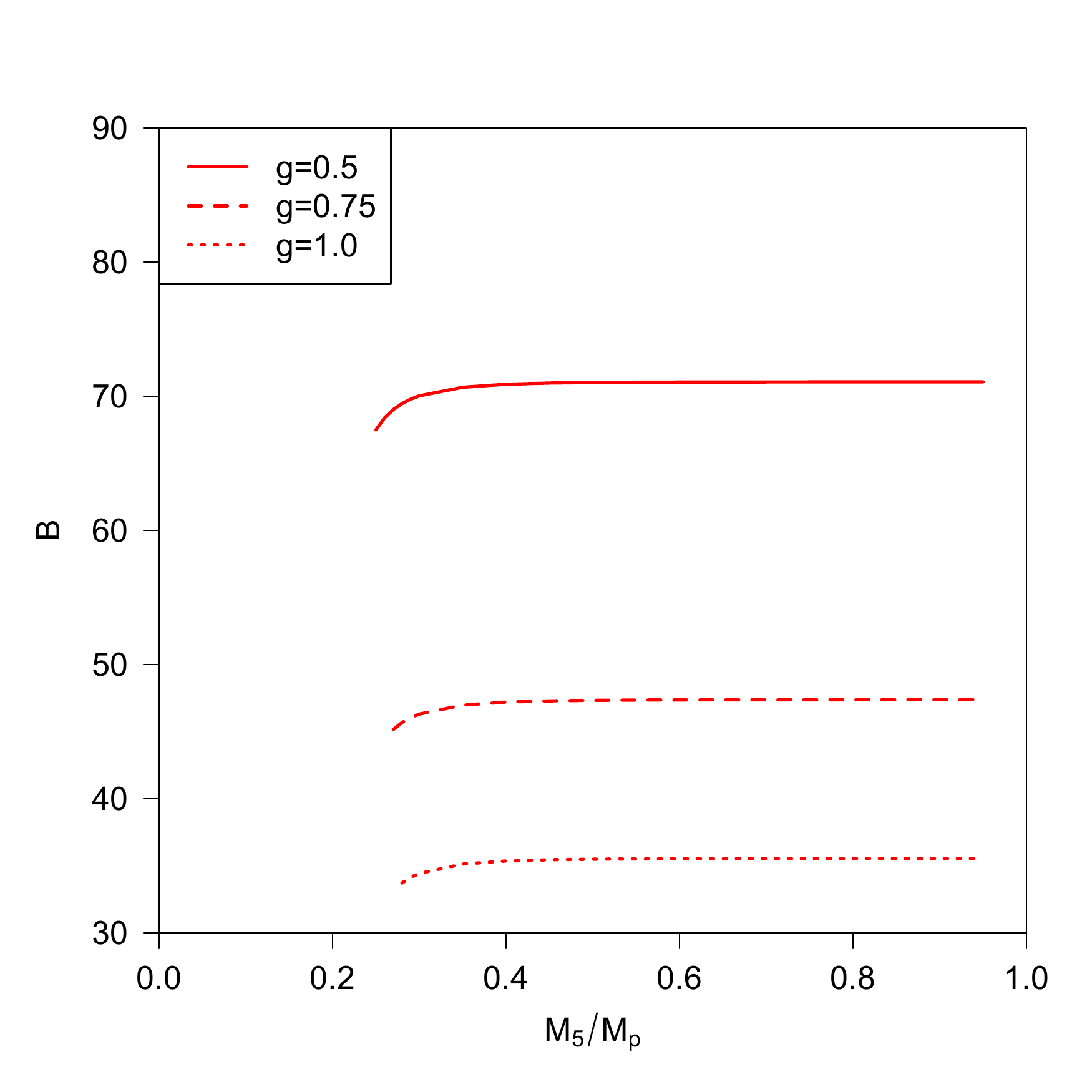}
\caption{The vacuum decay exponent $B$ for the quadratic potential 
plotted as a function of $M_5$ for barriers with $\phi_M=0.4M_p$ (left) 
and $\phi_M=0.1M_p$ (right). The exponent approaches the 4D value 
as $M_5$ approaches the 4D Planck mass $M_p$.}
\label{actionfig}
\end{figure}

Figure \ref{actionfig} shows the tunnelling exponent for the potential 
$V_q$ with the parameter sets considered in \S \ref{sec:setting},
these are plotted as a function of the mass parameter 
$M_5=M_p^{2/3}\ell^{-1/3}$,
which determines the strength of gravity in five dimensions.
The barrier is at $\phi_M=0.4M_p$ and $\phi_M=0.1M_p$. 
These test case examples show a reduction in $B$, hence an 
increase in the vacuum decay rate, due to the increasing influence
of the extra dimension.

The edge of the plots denotes a minimum value of $M_5$ beyond 
which the numerical solutions cease to exist.
Close to this limit, the total surface tension on the brane
becomes negative near the centre of the bubble. 
Note that the allowed range of $M_5$ is narrow, as in the examples plotted 
above, therefore does not correspond to a significant hierarchy.
Therefore adding an extra dimension
only affects the decay rate in very specialised situations. 

\begin{figure}
\centering
\includegraphics[width=0.5\linewidth]{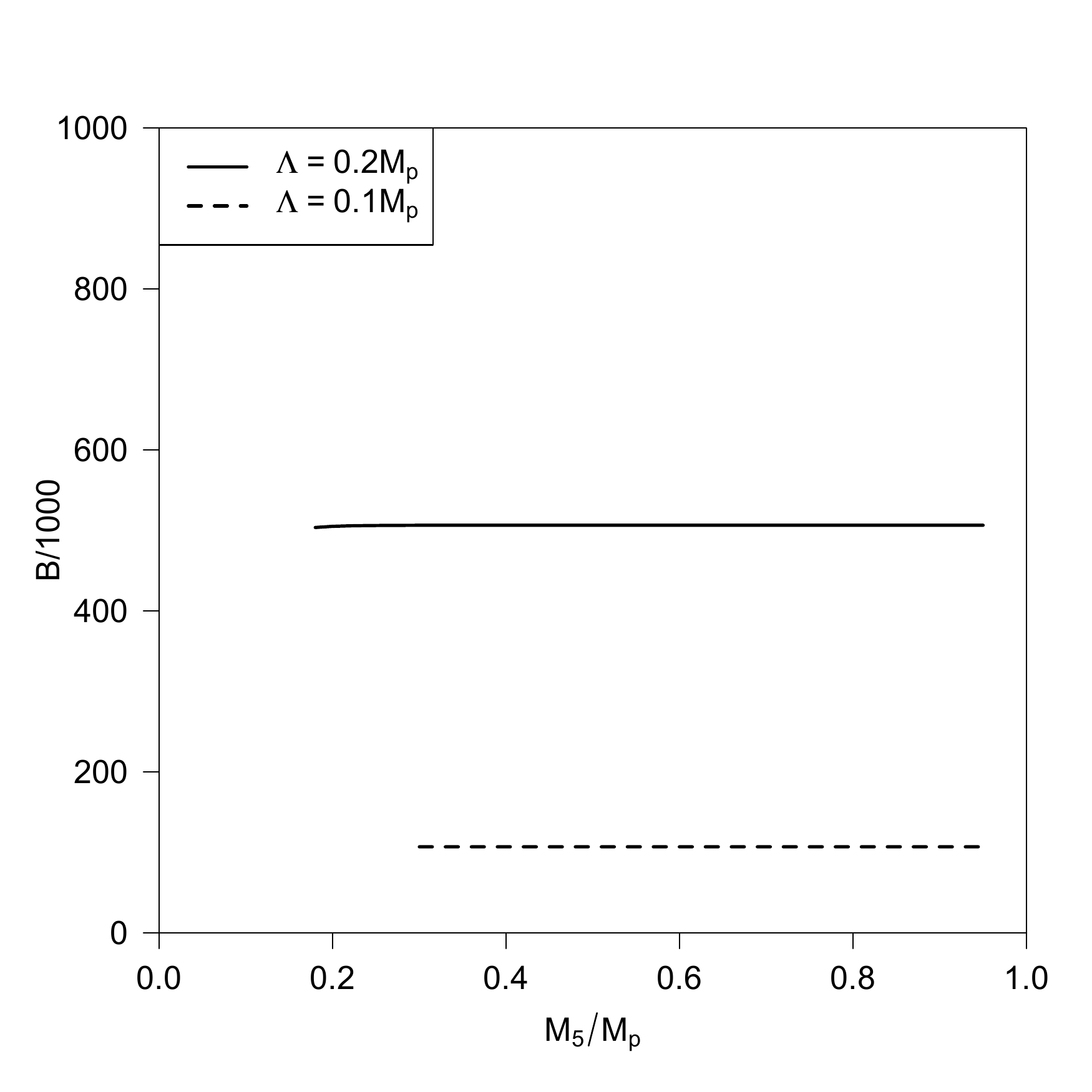}
\caption{The vacuum decay exponent $B$ plotted as a function of 
$M_5$ for Higgs potentials with a range of values or the instability 
scale $\Lambda$. There is no dependence on the extra dimension.}
\label{actionhiggs}
\end{figure}
We also show the tunnelling exponent for the Higgs-style potential
$V_H$, with parameters chosen within the Standard Model range 
in Fig. \ref{actionhiggs}.
The Higgs potential is small at the Planck scale because the parameter 
$g$ in the potential is small. Consequently, vacuum decay rates with 
the Higgs potential show no obvious dependence on the extra dimensions.

\section{Summary}
\label{sec:concl}

To sum up: we have found instanton solutions for a brane scalar
field representing vacuum decay from a critical RS flat brane. We
explored general bubble solutions, as well as an approximate Higgs 
potential. We calculated the tunnelling exponent for a range of
warping in the extra dimension, and compared it to that of a phase 
transition in 4D asymptotically flat space. The influence 
of the fifth dimension on tunnelling rates is relatively minor, 
except for a strongly backreacting bubble.

A Higgs-style potential was also considered, however, for realistic parameter
ranges, the impact of the extra dimension was negligible. This is to be
contrasted to the case of vacuum decay seeded by primordial brane
black holes, as in \cite{Cuspinera:2018woe}, where the decay rate is
significant. We conclude that, rather like the 4D case, black holes
are required to produce significant decay rates.

One interesting feature of our numerical solutions was that they had
a sharp cut-off in the allowed value of $M_5$, due to the brane tension 
becoming negative. This is possibly due to the fact we integrate out
from $\tau=a=0$, hence this method does not allow for a wormhole-type
solution where the brane transitions from positive to negative tension
as in \cite{Ida:2001qw,Ochiai:2001fh}. It might be interesting to consider
this further.


\acknowledgments

We are grateful for the hospitality of the Perimeter Institute, where part
of this research was undertaken. This work was supported in part by the 
Leverhulme grant \emph{Challenging the Standard Model with Black Holes}
and in part by STFC consolidated grant ST/P000371/1.  
LC acknowledges financial support from CONACyT,
RG is supported in part by the Perimeter Institute for Theoretical Physics,
and KM is supported by an STFC studentship.
Research at Perimeter Institute is supported by the Government of
Canada through the Department of Innovation, Science and Economic 
Development Canada and by the Province of Ontario through the
Ministry of Research, Innovation and Science.

\providecommand{\href}[2]{#2}
\begingroup\raggedright\endgroup


\begin{thebibliography}{10}


\bibitem{CDL}
S.~Coleman and F.~De~Luccia, 
{\it {Gravitational effects on and of vacuum decay}},  
Phys.Rev. {\bf D21} (1980) 3305--3315.

\bibitem{ATLAS:2012ae} 
G.~Aad {\it et al.}  [ATLAS Collaboration],
{\it Combined search for the Standard Model Higgs boson 
using up to 4.9 fb$^{-1}$ of $pp$ collision data at $\sqrt{s}=7$ 
TeV with the ATLAS detector at the LHC,}
Phys.Lett. {\bf B710}, 49 (2012)
[\href{http://arxiv.org/abs/1202.1408}{{\tt arXiv:1202.1408 [hep-ex]}}].
 
\bibitem{Chatrchyan:2012tx} 
S.~Chatrchyan {\it et al.}  [CMS Collaboration],
{\it Combined results of searches for the standard model 
Higgs boson in $pp$ collisions at $\sqrt{s}=7$ TeV,}
Phys.Lett. {\bf B710}, 26 (2012)
[\href{http://arxiv.org/abs/1202.1488}{{\tt arXiv:1202.1488 [hep-ex]}}].
  
\bibitem{Degrassi:2012ry} 
G.~Degrassi, S.~Di Vita, J.~Elias-Miro, J.~R.~Espinosa, 
G.~F.~Giudice, G.~Isidori and A.~Strumia,
{\it {Higgs mass and vacuum stability in the Standard Model at NNLO}},  
JHEP {\bf 1208} (2012) 098,
[\href{http://arxiv.org/abs/1205.6497}{{\tt arXiv:1205.6497 [hep-ph]}}].

\bibitem{Gorsky:2014una} 
A.~Gorsky, A.~Mironov, A.~Morozov and T.~N.~Tomaras,
{\it {Is the Standard Model saved asymptotically by conformal symmetry?}},  
JETP {\bf 120} (2015) 399-409,
[Zh.Eksp.Teor.Fiz. {\bf 147} (2015) 399-409]
[\href{http://arxiv.org/abs/1409.0492}{{\tt arXiv:1409.0492 [hep-ph]}}].

\bibitem{Bezrukov:2014ina} 
F.~Bezrukov and M.~Shaposhnikov,
{\it {Why should we care about the top quark Yukawa coupling?}},  
Zh.Eksp.Teor.Fiz. {\bf 147} (2015) 389,
[\href{http://arxiv.org/abs/1411.1923}{{\tt arXiv:1411.1923 [hep-ph]}}].

\bibitem{Ellis:2015dha} 
J.~Ellis,
{\it Discrete Glimpses of the Physics Landscape after the Higgs Discovery},
[\href{http://arxiv.org/abs/1501.05418}{{\tt arXiv:1501.05418 [hep-ph]}}].
 
\bibitem{Blum:2015rpa} 
K.~Blum, R.~T.~D'Agnolo and J.~Fan,
{\it {Vacuum stability bounds on Higgs coupling deviations}},  
[\href{http://arxiv.org/abs/1502.01045}{{\tt arXiv:1502.01045 [hep-ph]}}].

\bibitem{Krive:1976sg} 
I.~V.~Krive and A.~D.~Linde,
{\it {On the Vacuum Stability Problem in Gauge Theories}},  
Nucl.Phys. {\bf B432} (1976) 265.

\bibitem{1982Natur.298}
M.~S. Turner and F.~Wilczek,
{\it {Is our vacuum metastable}},  
Nature {\bf D79} (1982) 633.

\bibitem{Sher:1988mj} 
M.~Sher,
{\it Electroweak Higgs Potentials and Vacuum Stability,}
Phys.\ Rept.\  {\bf 179}, 273 (1989).

\bibitem{Isidori:2001bm} 
G.~Isidori, G.~Ridolfi and A.~Strumia,
{\it On the metastability of the standard model vacuum,}
Nucl.\ Phys.\ B {\bf 609}, 387 (2001)
[\href{http://arxiv.org/abs/hep-ph/0104016}{{\tt hep-ph/0104016}}].

\bibitem{coleman1977}
S.~Coleman, 
{\it {Fate of the false vacuum: Semiclassical theory}},  
Phys.Rev. {\bf D15} (1977) 2929--2936.

\bibitem{callan1977}
C. G. Callan and S.~Coleman,
{\it {Fate of the false vacuum II: First quantum corrections}},  
Phys.Rev. {\bf D16} (1977) 1762--1768.

\bibitem{Kobzarev:1974cp} 
I.~Y.~Kobzarev, L.~B.~Okun and M.~B.~Voloshin,
{\it {Bubbles in Metastable Vacuum}},  
Sov.J.Nucl.Phys. {\bf 20} (1975) 644,
[Yad.Fiz. {\bf 20} (1974) 1229].

\bibitem{GMW}
R.~Gregory, I.~G.~Moss and B.~Withers,
{\it {Black holes as bubble nucleation sites}},  
JHEP {\bf 1403} (2014) 081,
[\href{http://arxiv.org/abs/1401.0017}{{\tt arXiv:1401.0017 [hep-th]}}].

\bibitem{BGM1} 
P.~Burda, R.~Gregory and I.~Moss,
{\it Gravity and the stability of the Higgs vacuum},
Phys.\ Rev.\ Lett.\  {\bf 115}, 071303 (2015)
[\href{http://arxiv.org/abs/1501.04937}{{\tt arXiv:1501.024937 [hep-th]}}].

\bibitem{BGM2} 
P.~Burda, R.~Gregory and I.~Moss,
{\it Vacuum metastability with black holes},
JHEP {\bf 1508}, 114 (2015)
[\href{http://arxiv.org/abs/1503.07331}{{\tt arXiv:1503.07331 [hep-th]}}].

\bibitem{BGM3}
P.~Burda, R.~Gregory and I.~Moss,
{\it The fate of the Higgs vacuum},
JHEP {\bf 1606}, 025 (2016)
[\href{http://arxiv.org/abs/1601.02152}{{\tt arXiv:1601.02152 [hep-th]}}].

\bibitem{Gregory:2016xix} 
R.~Gregory and I.~G.~Moss,
{\it The Fate of the Higgs Vacuum},
PoS ICHEP {\bf 2016}, 344 (2016)
[\href{http://arxiv.org/abs/1611.04935}{{\tt arXiv:1611.04935 [hep-th]}}].

\bibitem{PhysRevD.35.1161}
W.~A. Hiscock,
{\it {Can black holes nucleate vacuum phase transitions?}},  
Phys.Rev. {\bf D35} (1987) 1161--1170.

\bibitem{Berezin:1987ea}
V.~Berezin, V.~Kuzmin, and I.~Tkachev, 
{\it {O(3) invariant tunneling in general relativity}},  
Phys.Lett. {\bf B207} (1988) 397.

\bibitem{Tetradis:2016vqb} 
N.~Tetradis,
{\it Black holes and Higgs stability},
JCAP {\bf 1609}, no. 09, 036 (2016)
[\href{http://arxiv.org/abs/1606.04018}{{\tt arXiv:1606.04018 [hep-ph]}}].

\bibitem{Chen:2017suz} 
P.~Chen, G.~Dom\`enech, M.~Sasaki and D.~h.~Yeom,
{\it Thermal activation of thin-shells in anti-de Sitter black hole spacetime},
JHEP {\bf 1707}, 134 (2017)
[\href{http://arxiv.org/abs/1704.04020}{{\tt arXiv:1704.04020 [gr-qc]}}].

\bibitem{Gorbunov:2017fhq} 
D.~Gorbunov, D.~Levkov and A.~Panin,
{\it Fatal youth of the Universe: black hole threat for the electroweak 
vacuum during preheating},
JCAP {\bf 1710}, no. 10, 016 (2017)
[\href{http://arxiv.org/abs/1704.05399}{{\tt arXiv:1704.05399 [astro-ph.CO]}}].

\bibitem{Mukaida:2017bgd} 
K.~Mukaida and M.~Yamada,
{\it False Vacuum Decay Catalyzed by Black Holes},
Phys.\ Rev.\ D {\bf 96}, no. 10, 103514 (2017)
[\href{http://arxiv.org/abs/1706.04523}{{\tt arXiv:1706.04523 [hep-th]}}].

\bibitem{ArkaniHamed:1998rs} 
N.~Arkani-Hamed, S.~Dimopoulos and G.~R.~Dvali,
{\it The Hierarchy problem and new dimensions at a millimeter},
Phys.Lett. {\bf B429}, 263 (1998) 263--272,
[\href{http://arxiv.org/abs/hep-ph/9803315}{{\tt hep-ph/9803315}}].

\bibitem{Antoniadis:1998ig} 
I.~Antoniadis, N.~Arkani-Hamed, S.~Dimopoulos and G.~R.~Dvali,
{\it New dimensions at a millimeter to a Fermi and superstrings at a TeV},
Phys.\ Lett.\ B {\bf 436}, 257 (1998)
[\href{http://arxiv.org/abs/hep-ph/9804398}{{\tt hep-ph/9804398}}].

\bibitem{Randall:1999ee} 
L.~Randall and R.~Sundrum,
{\it A Large mass hierarchy from a small extra dimension},
Phys.Rev.Lett.{\bf 83}, 3370 (1999) 3370--3373,
[\href{http://arxiv.org/abs/hep-ph/9905221}{{\tt hep-ph/9905221}}].

\bibitem{Randall:1999vf} 
L.~Randall and R.~Sundrum,
{\it An Alternative to compactification},
Phys.\ Rev.\ Lett.\  {\bf 83}, 4690 (1999)
[\href{http://arxiv.org/abs/hep-th/9906064}{{\tt hep-th/9906064}}].

\bibitem{Dimopoulos:2001hw}
S.~Dimopoulos and G.~L.~Landsberg,
{\it {Black holes at the LHC}},  
Phys.Rev.Lett. {\bf 87} (2001) 161602.
[\href{http://arxiv.org/abs/hep-ph/0106295}{{\tt hep-ph/0106295}}].

\bibitem{Giddings:2001bu} 
S.~B.~Giddings and S.~D.~Thomas,
{\it High-energy colliders as black hole factories: The End of short distance physics},
Phys.\ Rev.\ D {\bf 65}, 056010 (2002)
[\href{http://arxiv.org/abs/hep-ph/0106219}{{\tt hep-ph/0106219}}].

\bibitem{Landsberg:2003br} 
G.~L.~Landsberg,
{\it Black holes at future colliders and in cosmic rays},
Eur.\ Phys.\ J.\ C {\bf 33}, S927 (2004)
[\href{http://arxiv.org/abs/hep-ex/0310034}{{\tt hep-ex/0310034}}].

\bibitem{Cuspinera:2018woe} 
L.~Cuspinera, R.~Gregory, K.~Marshall and I.~G.~Moss,
{\it Higgs Vacuum Decay from Particle Collisions?},
Phys.\ Rev.\ D {\bf 99}, no. 2, 024046 (2019)
[\href{http://arxiv.org/abs/1803.02871}{{\tt arXiv:1803.02871 [hep-th]}}].  

\bibitem{Gregory:2001xu} 
R.~Gregory and A.~Padilla,
{\it Nested brane worlds and strong brane gravity},
Phys.\ Rev.\ D {\bf 65}, 084013 (2002)
[\href{http://arxiv.org/abs/hep-th/0104262}{{\tt hep-th/0104262}}].

\bibitem{Gregory:2001dn} 
R.~Gregory and A.~Padilla,
{\it Brane world instantons},
Class.\ Quant.\ Grav.\  {\bf 19}, 279 (2002)
[\href{http://arxiv.org/abs/hep-th/0107108}{{\tt hep-th/0107108}}].

\bibitem{Davis:2005hf} 
S.~C.~Davis and S.~Brechet,
{\it Vacuum decay on a brane world},
Phys.\ Rev.\ D {\bf 71}, 104023 (2005)
[\href{http://arxiv.org/abs/hep-ph/0503243}{{\tt hep-ph/0503243}}].

\bibitem{Demetrian:2005sr} 
M.~Demetrian,
{\it False vacuum decay in a brane world cosmological model},
Gen.\ Rel.\ Grav.\  {\bf 38}, 953 (2006)
[\href{http://arxiv.org/abs/gr-qc/0506028}{{\tt gr-qc/0506028}}].

\bibitem{Dudas:2007hq} 
E.~Dudas, J.~Mourad and F.~Nitti,
{\it Metastable vacua in brane worlds},
JHEP {\bf 0708}, 057 (2007)
[\href{http://arxiv.org/abs/0706.1269}{{\tt arXiv:0706.1269 [hep-th]}}].  

\bibitem{Ida:2001qw} 
D.~Ida, T.~Shiromizu and H.~Ochiai,
{\it Semiclassical instability of the brane world: Randall-Sundrum bubbles},
Phys.\ Rev.\ D {\bf 65}, 023504 (2002)
[\href{http://arxiv.org/abs/hep-th/0108056}{{\tt hep-th/0108056}}].

\bibitem{Ochiai:2001fh} 
H.~Ochiai, D.~Ida and T.~Shiromizu,
{\it Quantum creation of the Randall-Sundrum bubble},
Prog.\ Theor.\ Phys.\  {\bf 107}, 703 (2002)
[\href{http://arxiv.org/abs/hep-th/0111070}{{\tt hep-th/0111070}}].

\bibitem{Izumi:2007gs} 
K.~Izumi, K.~Koyama, O.~Pujolas and T.~Tanaka,
{\it Bubbles in the Self-Accelerating Universe},
Phys.\ Rev.\ D {\bf 76}, 104041 (2007)
[\href{http://arxiv.org/abs/0706.1980}{{\tt arXiv:0706.1980 [hep-th]}}].  

\bibitem{Sbisa:2014gwh} 
F.~Sbisà and K.~Koyama,
{\it Perturbations of Nested Branes With Induced Gravity},
JCAP {\bf 1406}, 029 (2014)
[\href{http://arxiv.org/abs/1404.0712}{{\tt arXiv:1404.0712 [hep-th]}}].  

\bibitem{Israel:1966}
W.~Israel,
{\it {Singular hypersurfaces and thin shells in general relativity}},  
Nuovo Cimento Soc. Ital. Phys. {\bf B44} (1966) 4349.

\bibitem{Chamblin:1999ya}
H.~A.~Chamblin and H.~S.~Reall,
{\it Dynamic dilatonic domain walls},
Nucl.\ Phys.\ B {\bf 562}, 133 (1999)
[\href{http://arxiv.org/abs/hep-th/9903225}{{\tt hep-th/9903225}}].

\bibitem{Kaloper:1999sm}
N.~Kaloper,
{\it Bent domain walls as braneworlds},
Phys.\ Rev.\ D {\bf 60}, 123506 (1999)
[\href{http://arxiv.org/abs/hep-th/9905210}{{\tt hep-th/9905210}}].

\bibitem{Kraus:1999it}
P.~Kraus,
{\it Dynamics of anti-de Sitter domain walls},
JHEP {\bf 9912}, 011 (1999)
[\href{http://arxiv.org/abs/hep-th/9910149}{{\tt hep-th/9910149}}].

\bibitem{Karch:2000gx}
A.~Karch and L.~Randall,
{\it Open and closed string interpretation of SUSY CFT's on branes with
boundaries},
JHEP {\bf 0106}, 063 (2001)
[\href{http://arxiv.org/abs/hep-th/0105132}{{\tt hep-th/0105132}}].

\bibitem{Binetruy:1999ut}
P.~Binetruy, C.~Deffayet and D.~Langlois,
{\it Non-conventional cosmology from a brane-universe},
Nucl.\ Phys.\ B {\bf 565}, 269 (2000)
[\href{http://arxiv.org/abs/hep-th/9905012}{{\tt hep-th/9905012}}].

\bibitem{BCG}
P.~Bowcock, C.~Charmousis and R.~Gregory,
{\it General brane cosmologies and their global spacetime structure},
Class.\ Quant.\ Grav.\  {\bf 17}, 4745 (2000)
[\href{http://arxiv.org/abs/hep-th/0007177}{{\tt hep-th/0007177}}].

\bibitem{Maartens:2000fg} 
R.~Maartens,
{\it Cosmological dynamics on the brane},
Phys.\ Rev.\ D {\bf 62}, 084023 (2000)
[\href{http://arxiv.org/abs/hep-th/0004166}{{\tt hep-th/0004166}}].

\bibitem{Garriga:1999yh} 
J.~Garriga and T.~Tanaka,
{\it Gravity in the brane world},
Phys.\ Rev.\ Lett.\  {\bf 84}, 2778 (2000)
[\href{http://arxiv.org/abs/hep-th/9911055}{{\tt hep-th/9911055}}].

\bibitem{Shiromizu:1999wj} 
T.~Shiromizu, K.~i.~Maeda and M.~Sasaki,
{\it The Einstein equation on the 3-brane world},
Phys.\ Rev.\ D {\bf 62}, 024012 (2000)
[\href{http://arxiv.org/abs/gr-qc/9910076}{{\tt gr-qc/9910076}}].

\bibitem{Sasaki:1999mi} 
M.~Sasaki, T.~Shiromizu and K.~i.~Maeda,
{\it Gravity, stability and energy conservation on the Randall-Sundrum brane world},
Phys.\ Rev.\ D {\bf 62}, 024008 (2000)
[\href{http://arxiv.org/abs/hep-th/9912233}{{\tt hep-th/9912233}}].


\end{thebibliography}
\end{document}